\documentclass{article}
\usepackage{amssymb}  
\usepackage{amsmath} 
\usepackage{authblk}
\usepackage{graphicx}
\usepackage{dcolumn}
\usepackage{bm}
\usepackage{color} 

\begin{document}

\newcommand{\CPS}{CoPS$_{3}$}
\newcommand{\MPS}{MnPS$_{3}$}
\newcommand{\NPS}{NiPS$_{3}$}
\newcommand{\FPS}{FePS$_{3}$}
\newcommand{\Cm}{\emph{C}2\emph{/m}}
\newcommand{\Pm}{\emph{P}1}
\newcommand{\signsi}{$\text{d}\sigma_{nsi}/\text{d}\Omega$}
\newcommand{\sigii}{$\text{d}\sigma_{ii}/\text{d}\Omega$}
\newcommand{\sigc}{$\text{d}\sigma_{coh}/\text{d}\Omega$}
\newcommand{\sigm}{$\text{d}\sigma_{mag}/\text{d}\Omega$}
\newcommand{\signsf}{$\text{d}\sigma_{NSF}/\text{d}\Omega$}
\newcommand{\sigsf}{$\text{d}\sigma_{SF}/\text{d}\Omega$}
\newcommand{\ILL}{Institut Laue-Langevin}
\newcommand{\Neel}{N\'eel}

\title{The magnetic exchange parameters and anisotropy of the quasi-two dimensional antiferromagnet NiPS$_3$}

\author[1,2,3]{D. Lan\c{c}on}
\affil[1]{Institut Laue-Langevin, CS 20156, 38042 Grenoble C\'edex 9, France}
\affil[2]{Ecole Polytechnique F\'ed\'erale de Lausanne, SB ICMP LQM, CH-1015 Lausanne, Switzerland}
\affil[3]{Laboratory for Neutron Scattering and Imaging, Paul Scherrer Institute WHGA/150, 5232 Villigen PSI, Switzerland}

\author[4]{R. A. Ewings}
\author[4]{T. Guidi}
\affil[4]{ISIS Pulsed Neutron and Muon Source, STFC Rutherford Appleton Laboratory, Harwell Campus, Didcot, OX11 0QX}

\author[5]{F. Formisano}
\affil[5]{Consiglio Nazionale delle Ricerche, Istituto Officina dei materiali, Operative Group in Grenoble, F-38042, Grenoble, France}

\author[1]{A. R. Wildes}
\affil[1]{Institut Laue-Langevin, CS 20156, 38042 Grenoble C\'edex 9, France}

\date{\today}
\maketitle

\begin{abstract}
{Neutron inelastic scattering has been used to measure the magnetic excitations in powdered {\NPS}, a quasi-two dimensional antiferromagnet with spin $S = 1$ on a honeycomb lattice.  The spectra show clear, dispersive magnons with a $\sim 7$ meV gap at the Brillouin zone center.  The data were fitted using a Heisenberg Hamiltonian with a single-ion anisotropy assuming no magnetic exchange between the honeycomb planes.  Magnetic exchange interactions up to the third intraplanar nearest-neighbour were required.  The fits show robustly that {\NPS} has an easy axis anisotropy with $\Delta = 0.3$ meV and that the third nearest-neighbour has a strong antiferromagnetic exchange of $J_3 = -6.90$ meV.  The data can be fitted reasonably well with either $J_1 < 0$ or $J_1 > 0$, however the best quantitative agreement with high-resolution data indicate that the nearest-neighbour interaction is ferromagnetic with $J_1 = 1.9$ meV and that the second nearest-neighbour exchange is small and antiferromagnetic with $J_2 = -0.1$ meV.  The dispersion has a minimum in the Brillouin zone corner that is slightly larger than that at the Brillouin zone center, indicating that the magnetic structure of {\NPS} is close to being unstable.}
\end{abstract}

%
\section{Introduction}

{\NPS} belongs to a family of quasi-two dimensional antiferromagnets \cite{Brec, Grasso}.  The family have layered structures with the $2+$ transition metal ions forming a honeycomb lattice in the $ab$ planes.  The compounds in the family are isostructural, all having the monoclinic space group $C\frac{2}{m}$ \cite{Ouvrard85}, and the $ab$ planes are weakly bound by van der Waals forces.  

The compounds show a variety of physical properties that make them interesting.  Other elements and molecules can be intercalated between the planes and the compounds have been extensively studied as potential battery materials \cite{Grasso}.  The compounds are Mott insulators, however recent experiments show that they can become metallic under an applied pressure \cite{Haines, WangY}, offering insight into electronic band theory and potentially into high-temperature superconductivity.  Individual layers can be delaminated, attracting the interest of the graphene community \cite{Park,Susner,WangF}.

They are also good model systems for testing the theory of magnetism in low dimensions.  Other members of the family include {\MPS}, which is a good example of a Heisenberg system \cite{Joy92,Wildes98,Wildes06}, and {\FPS}, which is a good example of an Ising system \cite{Joy92,Wildes12,Lancon}.  These compounds have been extensively studied for their model magnetic properties.  A less-studied member of the family is {\CPS}, which appears to have an XY-like anisotropy \cite{Wildes17}.  {\NPS} makes up the fourth member of the family.  Combined, the family represent an excellent platform for the study of magnetism on a two-dimensional honeycomb lattice, with spin $S$ = 5/2, 2, 3/2 and 1 for {\MPS}, {\FPS}, {\CPS} and {\NPS} respectively.

\begin{figure}
\includegraphics[width=3in]{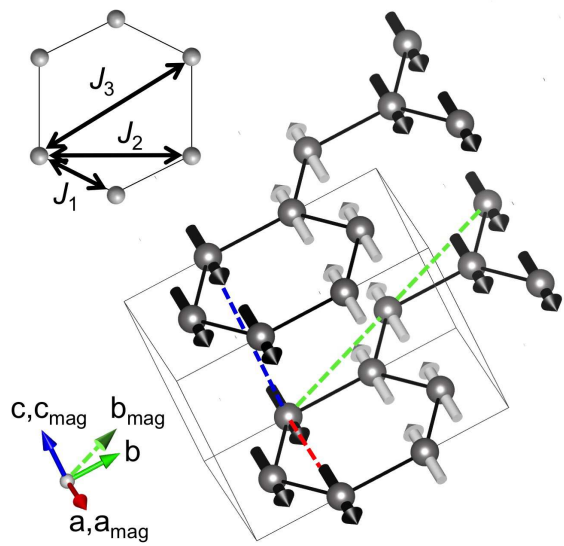}
\caption{\label{fig:MagStruc} The magnetic structure of {\NPS} with the crystallographic unit cell, and the unit cell used in the calculation of the magnetic dynamic structure factor. The insert shows the exchange interactions between the first, second and third nearest intraplanar neighbours.  The figure was created using the VESTA program \cite{VESTA}.}
\end{figure}

{\NPS} has the highest N{\'e}el temperature of the family with $T_N = 155$ K, forming the antiferromagnetic structure shown in figure \ref{fig:MagStruc} \cite{Wildes15}. The magnetic structure has a propagation vector of ${\bf{k}}_M = \left[010\right]$, forming zig-zag ferromagnetic chains parallel to the crystallographic $\bf{a}$ axis that are antiferromagnetically coupled along the $\bf{b}$ axis and ferromagnetically coupled along the $\bf{c}$ axis.  The moments are collinear with their common axis being almost parallel to $\bf{a}$.  

The magnetic susceptibility has a very broad maximum at $\sim 270$ K, well above the N{\'e}el temperature \cite{Joy92, Chandra, Wildes15}, which is a common feature of low-dimensional magnets \cite{deJongh}.  The susceptibility only becomes Curie-Weiss-like above $\sim 450$ K, indicating that critical fluctuations are very strong in this compound.  The data suggest that {\NPS} is a good example of a two-dimensional magnet.  

The paramagnetic susceptibility of {\NPS} has a large, negative Curie constant, $\Theta$ \cite{Joy92}.  A correlated effective field model has been used to analyse the susceptibility to determine a nearest-neighbour antiferromagnetic exchange interaction of $J_1 = -5.0$ meV and an easy-plane single-ion anisotropy of $\Delta = -1.39$ meV \cite{Chandra}.  There is some debate as to the nature of the anisotropy.  Initial measurements showed that the paramagnetic susceptibility was anisotropic \cite{Joy92, Chandra}, while more recent measurements showed it to be isotropic \cite{Wildes15}.  The discrepancy was attributed to the handling of the samples, with the act of gluing a sample to a support shown to affect the magnitude of the susceptibility \cite{Wildes15}.  This dependence, potentially linked to some form of magnetostriction or deformation of the sample, suggests that {\NPS} may be close to a magnetic instability.

Neutron inelastic scattering has previously been used to determine the magnetic exchange parameters of {\MPS} \cite{Wildes98} and {\FPS} \cite{Wildes12,Lancon}.  The technique gives direct access to the dynamic structure factor, $S\left({\bf{Q}},E\right)$, hence allowing the Hamiltonian to be tested and parameterised.  In this article, we report neutron inelastic scattering experiments on powdered samples of {\NPS}.  Estimates for the magnetic exchange parameters and anisotropy have been determined and are compared in a consistent manner with those for {\MPS} and {\FPS}.  The experiments and analysis closely follow those previously reported for powdered {\FPS} \cite{Wildes12}.

\section{Experiments}

Crystal samples of \NPS ~were grown by a vapour transport method using protocols that have been previously explained in detail \cite{Wildes15}.  Approximately 10 grams of crystals were ground to a powder.  The powdered sample was divided into three portions of approximately equal mass and each potion was compressed into a cylindrical pellet of 10 mm diameter.  The three pellets were placed side by side in an aluminium envelope with their cylindrical axes being collinear.

Neutron inelastic scattering measurements were performed using the MARI \cite{MARI} and MAPS \cite{MAPS} spectrometers at the ISIS facility, Rutherford Appleton Laboratories, UK, and using the BRISP spectrometer \cite{BRISP} at the Institut Laue Langevin, Grenoble.  These are all direct geometry spectrometers, using a fixed incident neutron energy, $E_i$, and measuring the neutron time-of-flight to determine the final neutron energy.  

MARI was used to give an overview of the magnetic excitations.  Measurements were performed with incident energies $E_i =$ 15, 30, 110 and 200 meV.  MAPS has a longer sample-detector path length than MARI and therefore has better energy resolution for the same incident energy.  It was used with $E_i =$ 200 meV to study in detail the scattering at small momentum transfers and large energy transfers.  BRISP is optimized for spectroscopic measurements at small scattering angles, and it was used to characterize a possible spin wave gap.  Measurements were performed with $E_i =$ 20.45 and 81.81 meV.  

The sample temperature was controlled using a closed-cycle cryorefrigerator for the ISIS spectrometers, and a liquid helium cryostat for the BRISP spectrometer.  The measurements were performed at the lowest possible temperature for the sample environment, which was 5 K for the cryorefrigerators and 1.5 K for the cryostat.

\section{\label{sec:DataAnalysis}Data Modelling and Analysis}

The MARI and MAPS data were reduced using the MANTID software suite \cite{MANTID}.  The LAMP software package was used to reduce the BRISP data \cite{LAMP}.  The data reduction involved normalizing to the incident flux, binning the data in rings with equivalent scattering angle, $\phi$, subtracting a background estimated from a measurement of the empty cryostat, and a normalization of the detector efficiency from a measurement of a vanadium standard.

The MARI and MAPS spectrometers have a large detector coverage, measuring the scattering to large neutron momentum transfers, $Q$.  The phonon contribution was estimated through the $Q-$dependence of scattering following a protocol described in the appendix.  The estimated phonon contribution was then subtracted from the data and the results were taken to be the magnetic inelastic scattering.

The magnetic inelastic scattering data were then modelled and fitted using linear spin wave theory.  The dynamic structure factor, $S\left({\bf{Q}},E\right)$, used to fit the data was derived from a Heisenberg Hamiltonian with a single-ion anisotropy:
\begin{equation}
\label{eq:Hamiltonian}
H=-\sum_{i,j}{J_{i,j}{\bf{S}}_i\cdot{\bf{S}}_j}-\Delta\sum_i{(S^z_i)^2},
\end{equation}
where $\Delta$ is the strength of the anisotropy and $J_{i,j}$ are the exchange interactions, with ferromagnetic exchange interactions being positive and antiferromagnetic exchange is negative. The same Hamiltonian was successfully used to model the magnon spectra for {\MPS} \cite{Wildes98} and {\FPS} \cite{Wildes12, Lancon}, and was used to estimate the magnetic exchange and anisotropy from the magnetic susceptibility of {\NPS} \cite{Chandra}.  

The crystal structure of {\NPS} is quoted to have some site disorder between the main $4g$ and the minority $2a$ sites for the Ni, and likewise for the main $4i$ and the minority $8j$ sites for the P \cite{Ouvrard85}.  However, it is likely that the minority contribution may be an artefact of the sample having stacking faults and refinements of the magnetic structure were not improved on including the site disorder \cite{Wildes15}.  Consequently, only the magnetic structure of the majority sites was considered in the analysis.

In keeping with previous calculations for {\FPS} \cite{Lancon}, $S\left({\bf{Q}},E\right)$ was derived from equation \ref{eq:Hamiltonian} by decomposing the antiferromagnetic structure of {\NPS} into four interlocking magnetic sublattices.  The sublattice vectors were chosen to be slightly different to the lattice vectors for the crystallographic unit cell.  Figure \ref{fig:MagStruc} shows the axes chosen for the calculation, with the subscript $mag$ designating the axes for a primitive sublattice.  The vectors {\bf{a}} = {\bf{a}$_{mag}$} and {\bf{c}} = {\bf{c}$_{mag}$}, however the vectors {\bf{b}} and {\bf{b}$_{mag}$} differ.  In the magnetic coordinates, $\left|{\bf{b}}_{mag}\right| = 2\left|{\bf{a}}\right|$ and $\gamma_{mag} = 120^{\circ}$.  The Miller indices for the two lattices are related through the transformation:
\begin{equation}
	\left[ \begin{array}{c} h \\ k \\ l \end{array}\right] = \left[\begin{array}{rcc} 1 & 0 & 0 \\ 1 & 1 & 0 \\ 0 & 0 & 1\end{array}\right]\left[ \begin{array}{c} h_{mag} \\ k_{mag} \\ l_{mag} \end{array}\right].
	\label{eq:TransMat}
\end{equation}

The propagation vector for {\NPS} is ${\bf{k}}_{M} = \left[010\right]$ while  it is ${\bf{k}}_{M} = \left[01\frac{1}{2}\right]$ for {\FPS} \cite{Lancon}.  Consequently, the transformation matrix given in equation \ref{eq:TransMat} is slightly different between the two compounds \cite{Lancon}.  Furthermore, while the matrix form of the Hamiltonian is identical between {\NPS} and {\FPS}, the matrix elements are slightly different.  After applying the Holstein-Primakoff transformations, the Hamiltonian for {\NPS} with its matrix elements are written:

\begin{equation}
\begin{array}{ll}
{\bf{H}_M}& = 2S\left[
\begin{array}{llll}
A & B^* & C & D^* \\
B & A     & D & C \\
C & D^* & A  & B^* \\
D & C    & B  & A\end{array}
\right] \\
A & =  2J_2\cos\left( 2\pi h_{mag}\right) +2J^{\prime}\cos\left(2\pi l_{mag}\right) \\ 
    & ~~~~~~-\Delta - J_1 +2 J_2 + 3J_3 - 4J^{\prime}  \\
B & = \exp\left(\frac{2\pi i}{3}\left[2h_{mag}+\frac{k_{mag}}{2}\right]\right) \\
    & \times\left\{
    \begin{array}{l}
     J_1\left(1+\exp\left(-2\pi ih_{mag}\right)\right) \\
    +J^{\prime}\left(
      \begin{array}{l}
           \exp\left(2\pi i l_{mag}\right) \\
           +\exp\left(-2\pi i \left[h_{mag} + l_{mag}\right]\right) 
      \end{array}
    \right)      
    \end{array}
    \right\} \\
C & = 2J_2\left(\cos\left(\pi k_{mag}\right)+\cos\left(2\pi\left[h_{mag}+\frac{k_{mag}}{2}\right]\right)\right) \\
D & =\exp\left(\frac{2\pi i}{3}\left[2h_{mag}+\frac{k_{mag}}{2}\right]\right) \\
    &\times\left\{
    \begin{array}{l}
    J_1\exp\left(-2\pi i\left[h_{mag}+\frac{k_{mag}}{2}\right]\right) \\
   +J_3\left(
     \begin{array}{l}
     2\cos\left(\pi k_{mag}\right) \\
     + \exp\left(-2\pi i\left[ 2h_{mag}+\frac{k_{mag}}{2}\right]\right)
     \end{array}
   \right)
   \end{array}
   \right\} \\
\end{array}
\label{eqn:H}
\end{equation}
where $J_{1..3}$ are the exchange interactions between the first to third nearest neighbours in the $ab$ planes, and $J^{\prime}$ is the exchange between neighbours along the {\bf{c}} axis.  As suggested from the paramagnetic susceptibility, the intraplanar exchange is expected to be weak due to the two-dimensional nature of {\NPS} and $J^{\prime}$ was assumed to be zero in the analysis of the neutron scattering data.

The Hamiltonian matrix in equation \ref{eqn:H} was then diagonalized to determine the eigenvectors, which were then used to calculate the magnetic dynamic structure factor, $S\left({\bf{Q}},E\right)$, and consequently the partial differential neutron cross-section.  Explicit equations for the eigenvectors of equation \ref{eqn:H} are given by Wheeler \emph{et al.} \cite{Wheeler}.

The resulting neutron cross-sections were used to fit the data collected using the MARI and MAPS spectrometers.  The procedure was identical to that used for {\FPS} and has been previously discussed in detail \cite{Wildes12}.  Summarizing briefly, experimental data were selected over a range of neutron scattering angles, $\phi$, and energy transfers, $E$.  Powder-averaged cross-sections were calculated for each data point in the selected range and convoluted with the instrument resolution, estimated using the CHOP utility program \cite{CHOP}.  Both experimental data and calculation were then summed over the chosen $\phi$ range to give one-dimensional functions of the intensity, $I\left(E\right)$, that could be compared in the fitting.  Different ranges of $\left(\phi,E\right)$ were selected and fitted in order to test the uniqueness of the resulting best fit parameters.

\section{\label{sec:Results}Results}

\begin{figure*}
\includegraphics[width=6in]{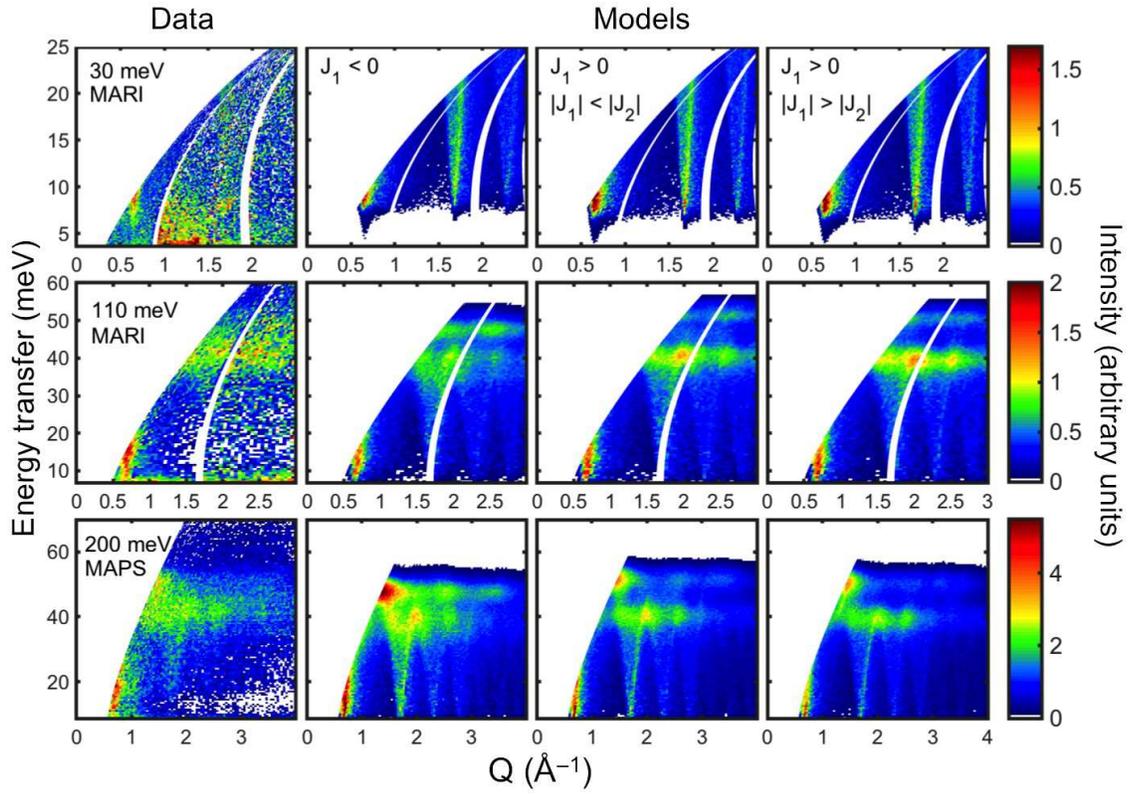}
\caption{\label{fig:Maps} Neutron inelastic data of {\NPS} measured on the MARI and MAPS spectrometers for a selection of different incident energies, along with the calculation for models with different fitted parameters.  The data were measured at a temperature of 5 K.}
\end{figure*}

The neutron inelastic scattering from the magnetic fluctuations in {\NPS} measured at 5 K are shown in figure \ref{fig:Maps}.  The figure shows data measured on MAPS and MARI for a selection of incident neutron energies, $E_i$.  The data have had estimates for the phonon contribution subtracted, following the procedure in the appendix, and are plotted from a non-zero minimum energy transfer, $E$, such that the strong elastic scattering is not visible.  

The data all show clear magnetic inelastic scattering which is particularly strong for $Q < 2$ {\AA}$^{-1}$.  The MARI data also showed some extra scattering, which is particularly visible for $E_i = 30$ meV within the range $1 \leq Q \leq 2$ {\AA}$^{-1}$ and $E \leq 8$ meV.  The position and relative intensity of the extra scattering depended on the choice of the incident neutron energy, showing that it was due to the instrument configuration and not representative of the sample.  

\begin{figure}
\includegraphics[width=3.5in]{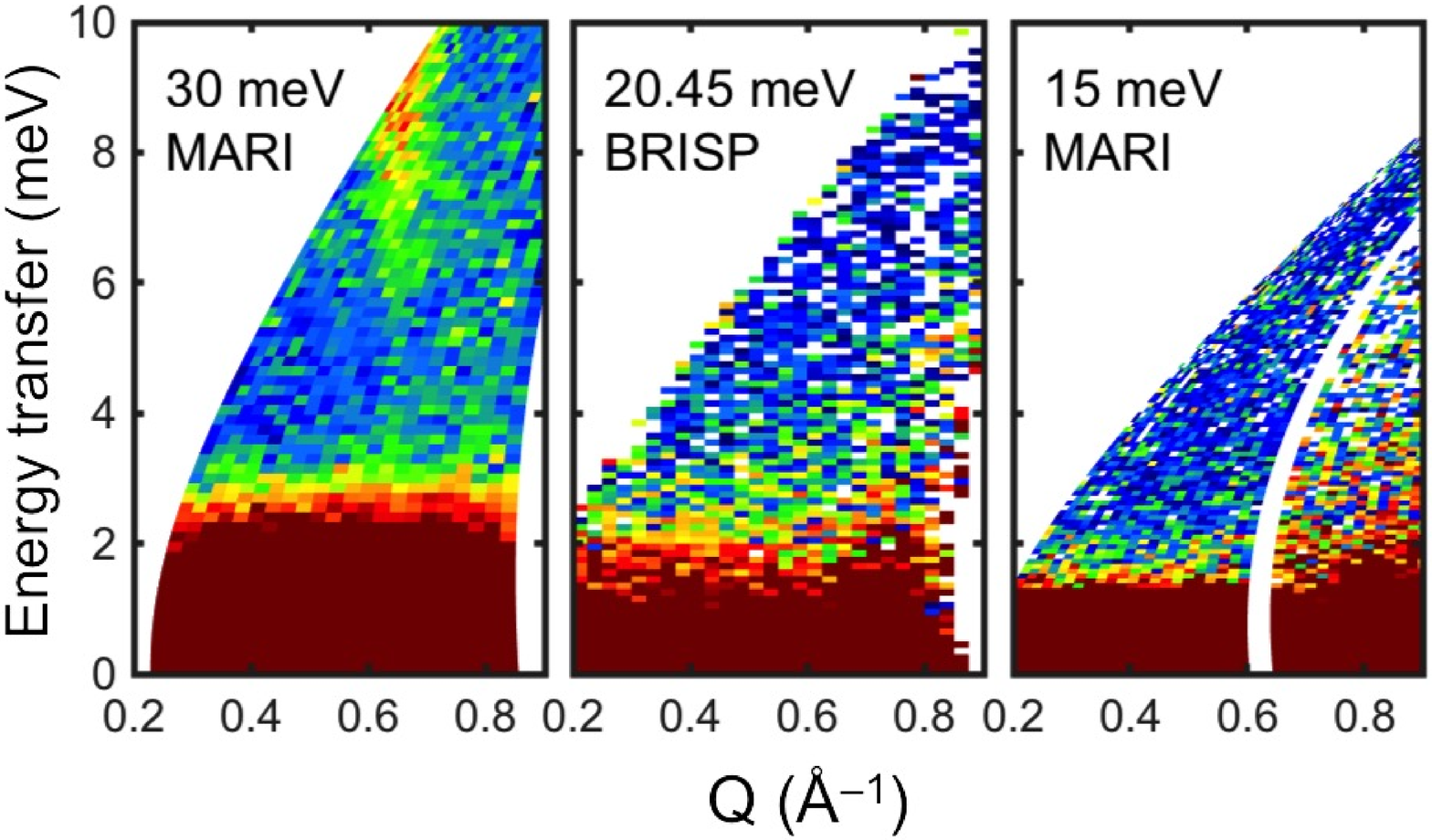}
\caption{\label{fig:BRISP} Neutron inelastic data for {\NPS} at low energy transfers measured on the BRISP and MARI spectrometers for incident neutron energies of 20.45 meV and 15 meV respectively.  The sample temperatures were 1.5 K on BRISP and 5 K on MARI.}
\end{figure}

Strong dispersive intensity is seen at $Q \sim 0.6$ {\AA}$^{-1}$ and small energy transfers.  Other, weaker, dispersive modes can be seen at $Q \sim 1.75$ {\AA}$^{-1}$ and, just visible in the MAPS data, at $\sim 2.4$ {\AA}$^{-1}$.  Neutron powder diffraction shows that these $Q$ points correspond to magnetic Bragg peaks, with the strongest peak being the $\left(010\right)$ at  $Q \sim 0.6$ {\AA}$^{-1}$ \cite{Wildes15}.  The magnetic scattering appears to have an energy gap at  this $Q$.  The gap is most clearly seen in the MARI data with $E_i = 30$ meV, which are shown on an expanded scale in figure \ref{fig:BRISP}.  The size of the gap is difficult to estimate precisely from these data, however measurements on BRISP and MARI with smaller $E_i$ allow a lower limit to be placed.  The energy and momentum transfers are coupled for neutrons, limiting the $\left(Q,E\right)$ range accessible for neutron scattering.  Measurements on BRISP, with $E_i = 20.45$ meV, and MARI, with $E_i = 15$ meV, limit the maximum measurable energy transfers at $Q = 0.6$ {\AA}$^{-1}$ of $7.3$ meV and $5.8$ meV for BRISP and MARI respectively.  These data are also shown in figure \ref{fig:BRISP}.  Neither data set shows any clear magnetic signal, suggesting that the gap must be $\gtrsim 7$ meV.  The presence of a spin wave gap establishes that {\NPS} has a finite magnetic anisotropy, $\Delta$.

\begin{figure}
\includegraphics[width=3.5in]{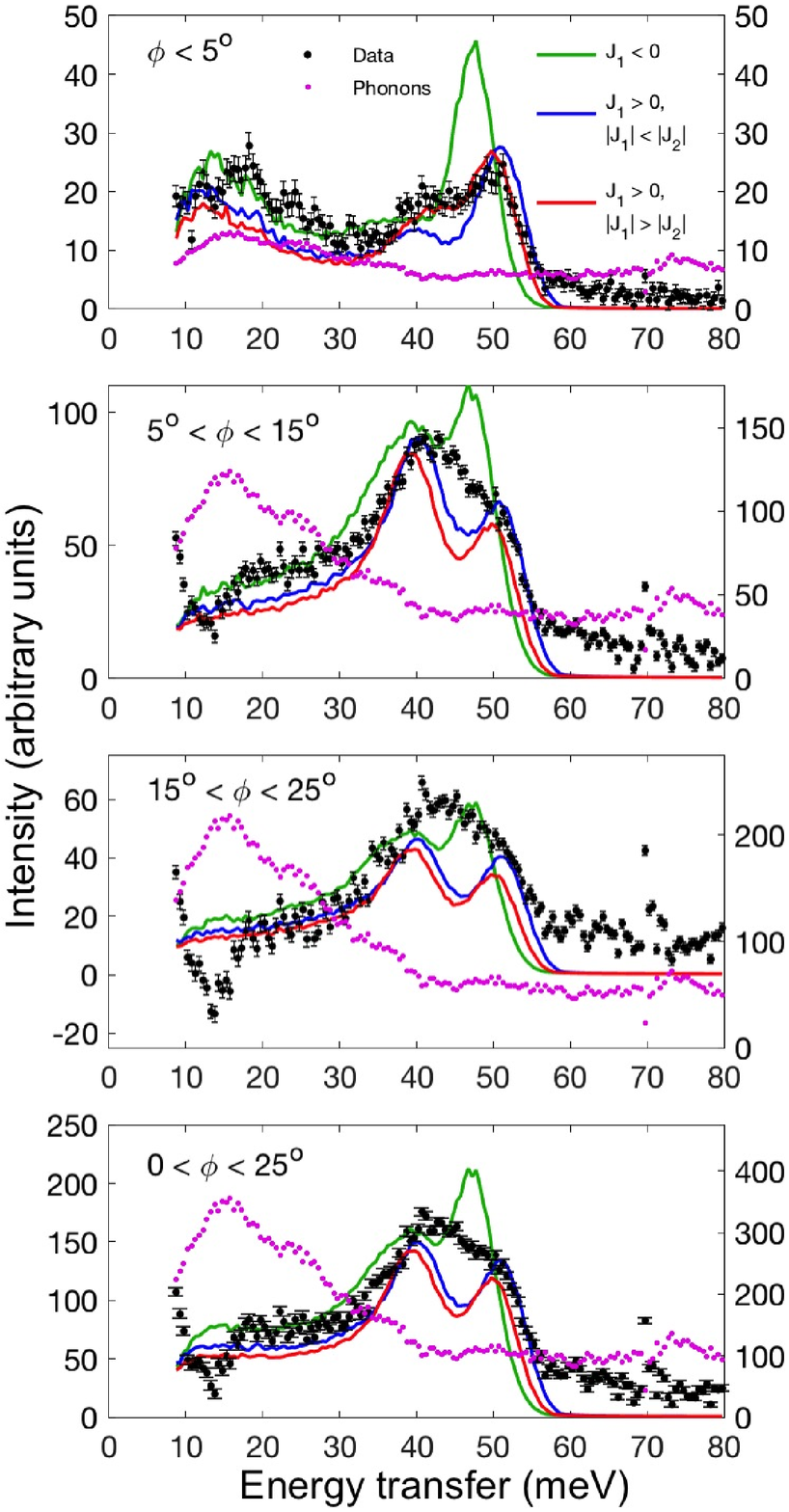}
\caption{\label{fig:Cuts} Magnetic $I\left(E\right)$ data, measured using MAPS with $E_i = 200$ meV, integrated over different ranges of the scattering angle, $\phi$.  Models with the exchange parameters from various fits are also shown, along with the estimated phonon contribution that had been subtracted from the total scattering.}
\end{figure}

Figure \ref{fig:Cuts} shows the MAPS data as a function of $E$, summed over various ranges of $\phi$.  The data have had the estimated phonon contribution subtracted and the contribution for each range, as determined by the method described in the appendix, is also shown in the figure.  The phonon contribution becomes large below $\sim 30$ meV, with a peak at $\sim 15$ meV.  The estimated magnetic contributions show a dip at approximately the same energy, with the data for $15^{\circ} < \phi < 25^{\circ}$ even showing negative intensities.  The phonon subtraction is notoriously difficult to get right at these energies, and the dip indicates that the phonon contribution is slightly overestimated in the $10 \leq E \leq 20$ meV energy range.  The overestimation is more problematic at larger scattering angles where the phonon contribution is stronger and the magnetic contribution is weaker.  For this reason, the fitting concentrated on the data for $\phi < 5^{\circ}$ where the influence of any phonon overestimation is minimalized.

The magnetic intensity shows substantial spectral weight from $35 \lesssim E \lesssim 55$ meV.  The spectral weight appears to form two broad bands: one centred at $E \sim 40$ meV and the other at $E \sim 50$ meV.   The bands are readily apparent in figure \ref{fig:Cuts}.  The spectral weight in the  $E \sim 50$ meV band is greater than for the $E \sim 40$ meV band at the smallest $\phi$, however the reverse is true for larger $\phi$.  This shift in the spectral weight between the two bands proved to be essential in determining the best estimate for the magnetic exchange interactions. 

Exchange interactions up to the third nearest-neighbour had to be included in the fits in order to have any reasonable comparison with the data.  The necessity of including $J_3$ in the fits was not unexpected, as this was also required to fit the spin wave dispersions of {\MPS} \cite{Wildes98} and {\FPS} \cite{Wildes12, Lancon}.   The values for $J_3$ proved to be very robust on fitting, consistently giving values of $\sim -6.5$ meV irrespective of the chosen range of $\left(\phi,E\right)$.  The fits establish $J_3$ to be large and antiferromagnetic and to be the dominant exchange in {\NPS}.

The fitted values for the anisotropy also proved to be robust, giving values of $\Delta \sim 0.3$ meV and establishing the single-ion anisotropy to have an easy axis.  This is in contrast to the analysis of the magnetic susceptibility, which concluded that the anisotropy is easy-plane \cite{Chandra}.  However, an easy-axis anisotropy gives a more logical description as the moments are collinear in the ordered magnetic structure, and are not coplanar as they have a small component out of the $ab$ planes \cite{Wildes15}.  Furthermore, the sign of $\Delta$ was a free parameter in the fits and calculations for $\Delta < 0$ did not give stable solutions.

Determining values for $J_1$ and $J_2$ from the fits proved to be more ambiguous.  Previous analysis of the magnetic susceptibility gave an exchange of $-5.0$ meV \cite{Chandra}, i.e. an antiferromagnetic exchange, however this estimate reflects the average exchange over all nearest neighbours.  Stability phase diagrams have been generated for the magnetic structures on a honeycomb lattice with up to three nearest-neighbours \cite{Rastelli, Fouet}.  For appropriate ratios of $J_2/J_1$ and $J_3/J_1$, the magnetic structure for {\NPS} is stable for either sign of $J_1$.  The MARI data could be fitted equally well with either a positive or negative $J_1$.  The values for $J_2$ would change accordingly, with a relation that empirically appeared to be $J_1-J_2 \approx 2$ meV.  

The ambiguity was lifted on close inspection of fits to the MAPS data.  MAPS has significantly better energy resolution that MARI, and measurements using a higher incident neutron energy gave access to high energies at smaller $Q$.   Figure \ref{fig:Cuts} shows a selection of fits to the data.  The resulting parameters are given in table \ref{tab:ExchangeVals}, and calculations of the expected magnetic scattering using these parameters are also shown in figure \ref{fig:Maps}.  All the parameters in the table are consistent with the magnetic structure of {\NPS}, as given by the calculated stability phase diagrams \cite{Rastelli, Fouet}.  

All the models give two peaks in the intensity from $35 \lesssim E \lesssim 55$ meV at $\phi > 5^{\circ}$.  The MARI data had insufficient resolution to differentiate the spectral weight in each of the peaks.  However, they are more clearly seen in the MAPS data and it is clear that their spectral weights are best fitted by models with $J_1 > 0$, i.e. a ferromagnetic exchange.  The conclusion becomes more apparent when comparing the data for $\phi < 5^{\circ}$, where the shift in spectral weight between the two peaks is reproduced for $J_1 > 0$ while only one clear peak is seen for fits with $J_1 < 0$.  Figure \ref{fig:Maps} also shows the calculated scattering for the model parameters in table \ref{tab:ExchangeVals}.  A qualitative inspection shows that the models with $J_1 > 0$ better resemble the measured data.

\begin{table}
\caption{Table showing the fit parameters used to calculate the expected neutron inelastic scattering in figures \ref{fig:Maps} and \ref{fig:Cuts}, and the magnon dispersions in figure \ref{fig:Spagetti}.  All values are in meV.}
\label{tab:ExchangeVals}
\begin{tabular}{ccccc}
 & $J_1 < 0$ & $J_1 > 0$                                    & $J_1 > 0$\\
 &                  & $\left|J_1\right| < \left|J_2\right|$ & $\left|J_1\right| > \left|J_2\right|$ \\
\hline
 $J_1$     & $-0.37$ &  $0.87$ &  $1.84$\\
 $J_2$    & $-1.98$  & $-1.38$ & $-0.18$\\
 $J_3$     & $-6.22$ & $-6.55$ & $-6.95$\\
 $\Delta$  &  $0.41$ &  $0.30$ &  $0.29$\\
 \hline
 $E_{\Gamma}$  & $9.44$ & $7.66$ & $6.79$\\
$E_{\text{C}}$     & $7.79$ & $7.04$ & $7.28$
\end{tabular}
\end{table}

 Two fits with $J_1 >0$ are shown in figure \ref{fig:Cuts}: one with $\left|J_1\right| < \left|J_2\right|$ and one with $\left|J_1\right| > \left|J_2\right|$.  The fits are practically identical if they are compared for $\phi > 5^{\circ}$.  This is also apparent when comparing the calculated intensities in figure \ref{fig:Maps}, with the two models being almost indistinguishable for the two sets of MARI data.  However, the fits with $\left|J_1\right| > \left|J_2\right|$ compare better with the data for $\phi < 5^{\circ}$.  This region was only accessible with sufficient resolution using MAPS.  
 
\begin{figure}
\includegraphics[width=3.5 in]{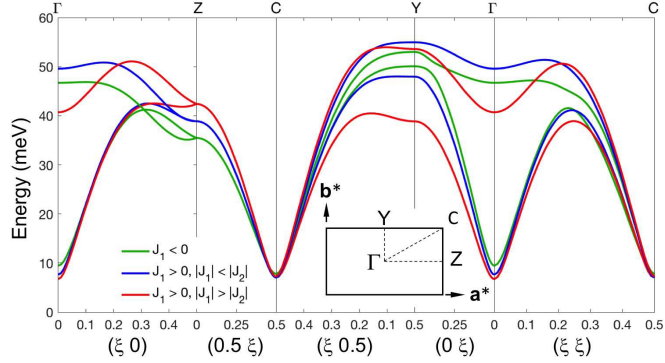}
\caption{\label{fig:Spagetti} The magnon dispersion along different trajectories in the Brillouin zone for the models listed in table \ref{tab:ExchangeVals}.  All the trajectories are given with respect to the crystallographic unit cell.  The Brillouin zone and the relevant positions are shown in the insert.}
\end{figure}

The preference is confirmed on comparing the calculated magnon energies for the different fits.  Figure \ref{fig:Spagetti} show the magnon dispersions for different trajectories around the Brillouin zone, calculated using the parameters in table \ref{tab:ExchangeVals}.  The dispersions show a number of common features.  All the dispersions have two doubly-degenerate magnon branches throughout the Brillouin zone, except at the Brillouin zone boundary between points Z and C where all the magnons are degenerate.
 
 The magnon energies at C, in the Brillouin zone corner, are of particular note.  All the calculations show a clear minimum at this point which is similar in magnitude to the minimum at the Brillouin zone centre.  For the magnetic structure to be stable, however, the minimum energy in the magnon dispersion must be at the Brillouin zone centre.  This consideration allows extra constraints to be placed on the exchange parameters.
 
 The magnon energies are given by the eigenvalues of equation \ref{eqn:H}, which take the form:
\begin{equation}
	\begin{array}{ll}
		\frac{E^2_{\bf{q}}}{4S^2} =& A^2+\left|B\right|^2-C^2-\left|D\right|^2 \\
		                                         & \pm\left(4\left|AB^* - CD^*\right|^2 - \left|BD^* -DB^*\right|\right)^{\frac{1}{2}}.
	\end{array}
\label{eq:Evals}
\end{equation}
The energy of the lowest magnons at the Brillouin zone centre is thus given by:
\begin{equation}
	E_{\Gamma} = 2S \left( \Delta \left( \Delta - 2J_1 - 8J_2 - 6J_3\right)\right)^{\frac{1}{2}},
\label{eq:E00}
\end{equation}
and the energy at C is given by:
\begin{equation}
	E_{\text{C}} = 2S \left( \Delta \left( \Delta + 2J_1 - 6J_3\right)\right)^{\frac{1}{2}}.
\label{eq:EC}
\end{equation}
Applying the condition $E_{\Gamma} < E_{\text{C}}$ leads to the inequality:
\begin{equation}
     J_1 > -2J_2.
\label{eq:Ineq}
\end{equation}
The calculated values for $E_{\Gamma}$ and $E_{\text{C}}$ are shown in table \ref{tab:ExchangeVals}.  All the parameters give an energy gap comparable to the lower limit suggested by figure \ref{fig:BRISP}, i.e. $E_{\Gamma} \gtrsim 7$ meV.   However, the inequality is respected only in the case of $J_1 > 0$, $\left|J_1\right| > \left|J_2\right|$.  Thus, $J_1$ is relatively large and positive and $J_2$ is relatively small and, most likely, negative.

Numerous fits were attempted with the added constraint of equation \ref{eq:Ineq}, including fixing $J_2 = 0$, over different ranges of $\left(\phi,E\right)$.  Fixing $J_2 = 0$ gave a fit result that was almost indistinguishable from the result for $J_1 > 0$, $\left|J_1\right| > \left|J_2\right|$ shown in figure \ref{fig:Cuts}.  While the errors on the parameters from an individual fit were typically in the second decimal place, the best estimate for the final values and their uncertainties comes from the spread in the fitted parameters over different fits.  The final parameters may be taken to be $J_1 = 1.9 \pm 0.1$ meV, $J_2 = -0.1 \pm 0.1$ meV, $J_3 = -6.90 \pm 0.05$ meV, and $\Delta = 0.3 \pm 0.1$ meV, giving energies of $E_{\Gamma} = 6.81$ meV and $E_{\text{C}} = 7.39$ meV.

 \section{\label{sec:Discussion}Discussion}

\begin{table}
\caption{Table showing the estimates for the anisotropies and exchange parameters, given in meV, for {\MPS} \cite{Wildes98},  {\FPS} \cite{Lancon}, and {\NPS}.}
\label{tab:ExchangeComp}
\begin{tabular}{cccc}
                     & {\MPS}                        & {\FPS}                         & {\NPS}\\
\hline
$S$               & $5/2$                          & $2$                             & $1$  \\
$T_N$	     & $78$ K                       & $120$ K                      & $155$ K \\
\hline
 $J_1$           &  $-0.77\left(9\right)$   &  $~~1.46\left(1\right)$ & $~~1.9\left(1\right)$ \\
 $J_2$           & $-0.07\left(7\right)$    & $-0.04\left(4\right)$     & $-0.1\left(1\right)$ \\
 $J_3$           & $-0.18\left(1\right)$    & $-0.96\left(5\right)$     & $-6.90\left(5\right)$\\
 $J^{\prime}$ & $0.0019\left(2\right)$ & $-0.0073\left(3\right)$ & $-$ \\
 $\Delta$ &  $0.0086\left(9\right)$ &  $~~2.66\left(8\right)$      &  $~~0.3\left(1\right)$\\
 \end{tabular}
\end{table}

The best estimates for the magnetic exchange parameters and the anisotropy of {\NPS} are listed in in table \ref{tab:ExchangeComp}.  Noting that a honeycomb lattice has three first nearest-neighbours, six second nearest-neighbours and three third nearest-neighbours, the weighted sum of these parameters is $-5.2$ meV which compares favourably with $-5.0$ meV, being the average value of the exchange determined from the analysis of the magnetic susceptibility \cite{Chandra} .  The exchange parameters can also be used to estimate the N{\'e}el, $T_N$, and Curie-Weiss, $\Theta$, temperatures for {\NPS}.  Mean field theory gives the following relations:   
\begin{equation}
	\begin{array}{ll}
		k_B\Theta &=\frac{2}{3}S\left(S+1\right)\left(3J_1+6J_2+3J_3\right) \\
		k_BT_N     &=\frac{2}{3}S\left(S+1\right)\left(J_1-2J_2-3J_3\right).
	\end{array}
\label{eq:TnTheta}
\end{equation}
Substituting the values from table \ref{tab:ExchangeComp} gives $\Theta = -241$ K and $T_N = 353$ K.  The calculated Curie-Weiss temperature is remarkably close to previously published values of $\Theta = -241$ and $\Theta = -254$ K \cite{Joy92}, providing confidence that the estimates for the exchange parameters are broadly correct.  The  calculated N{\'e}el temperature is more than twice the measured $T_N$, however this is often the case for compounds that exhibit strong critical fluctuations where mean field theory will break down.  A similar difference was observed for {\MPS} \cite{Wildes98}, where critical fluctuations are very strong \cite{Wildes06}.  The broad maximum in the susceptibility for {\NPS} \cite{Wildes15} is also seen in {\MPS} \cite{Joy92}, hence critical fluctuations are also likely to be very strong in the nickel compound.

Table \ref{tab:ExchangeComp} also lists the magnetic exchange parameters and anisotropy for {\MPS} and {\FPS}.  A comparison of the values for the three compounds shows an interesting evolution of the exchange parameters with the spin on the $2+$ transition metal ion.  The magnitudes of all the exchanges but $J_2$ increase with decreasing spin, which is also reflected in the magnitudes of the N{\'e}el temperatures.   All the compounds are antiferromagnets, but only {\MPS} has a nearest-neighbour exchange that is antiferromagnetic, i.e. negative.  Such an exchange is consistent with the ${\bf{k}}_M = 0$ magnetic structure of {\MPS}, with each magnetic moment antiferromagnetically coupled with all three of its nearest neighbours.  $J_1$ is positive, and therefore ferromagnetic, for {\FPS} and {\NPS}.  Their magnetic structures are stabilised by the strong antiferromagnetic third nearest-neighbour exchanges.  $J_3$ is particularly strong for {\NPS} where it is the dominant exchange.  The values for $J_2$ are close to zero for all the compounds.

\begin{figure}
\includegraphics[width=2.5 in]{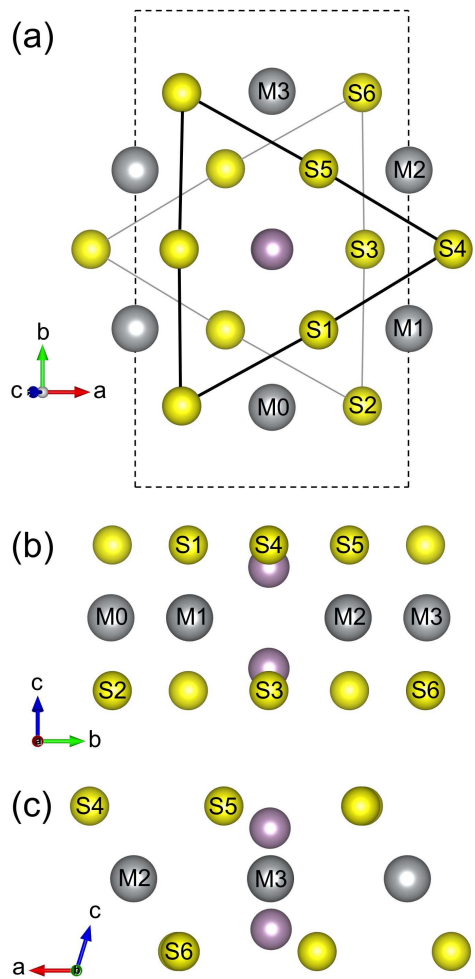}
\caption{\label{fig:Paths} Schematic showing the atoms close to the $ab$ plane for transmission metal-PS$_3$ compounds.  The transition metal atoms are shown in grey, the sulfur atoms in yellow and the phosphorus in purple. (a) shows the projection when viewing along ${\bf{c}}^{\star}$, with sulfur atoms above the $ab$ plane connected with black lines and sulfur atoms below the plane connected with grey lines.  The dashed line shows the size of the monoclinic unit cell in the $ab$ plane.  Selected metal atoms are marked as $M0..M2$, and selected sulfur atoms are marked as $S1..S6$.  (b) and (c) show views along the ${\bf{a}}$ and ${\bf{b}}$ axes respectively, with the same atoms marked.  Note that some of the marked atoms will hide others when viewed along certain axes.  The figure was created using the VESTA program \cite{VESTA}.}
\end{figure}

The values for the exchange parameters will be due to the possible superexchange pathways and to the magnitude of the orbital angular momentum of the $2+$ transition metal atoms.  A detailed calculation of these pathways will be the subject of future work, however some qualitative observations will be made here.  Figure \ref{fig:Paths} shows part of the crystal structure for the transmission metal-PS$_3$ compounds \cite{Ouvrard85}, showing those atoms closest to the $ab$ planes.  Selected transmission metal atoms are marked with $M$ while selected sulfur atoms are marked with $S$.  Each $M$ atoms has an approximately octahedral coordination with its neighbouring S atoms.

The nearest-neighbour exchange, $J_1$, is likely to be mediated through superexchange with sulfur atoms.  There are two S atoms on edge-shared octahedra between neighbouring $M$ atoms, for example $S1$ and $S2$ between atoms $M0$ and $M1$ in figures \ref{fig:Paths}(a) and (b).  The $M0-S-M1$ angle is $\sim 85^{\circ}$.  The Goodenough-Kanamori rules suggest that this interaction for Ni$^{2+}-$Ni$^{2+}$ and Fe$^{2+}-$Fe$^{2+}$ should be ferromagnetic \cite{Kanamori}, as is observed.  Mn$^{2+}$ has a half-filled d-shell and nominally has no orbital momentum, which appears to switch the sign of $J_1$ to be antiferromagnetic.

There is no easy superexchange route for second nearest neighbours.  As shown in figure \ref{fig:Paths}, the path between $M0$ and $M2$ would need to pass through two S atoms.  The path through $S1$ and $S3$ is highly unlikely because, as can be seen in figure \ref{fig:Paths}(b), $S1$ is above the $ab$ plane and $S3$ is below the plane.  Paths along  $S1 - S4$  and $S1 - S5$ are also unlikely.  While these atoms are above the $ab$ plane, the $M0-S1-S\left(4,5\right)-M2$ paths are not coplanar implying that non-overlapping orbitals would need to be involved.  The lack of a superexchange pathway would explain why $J_2$ is close to zero for all the compounds in table \ref{tab:ExchangeComp}.

A superexchange pathway is available for third nearest-neighbours.  Figures \ref{fig:Paths}(a) and (c) show that the $M0-S1-S5-M3$ path is coplanar and involves two atoms above the $ab$ plane.  This exchange pathway would need to be antiferromagnetic, and to increase in strength with decreasing spin and correspondingly changing orbital angular momentum.

A comparison of the anisotropies, $\Delta$, is also interesting.   {\MPS} has a very small anisotropy, most likely dominated by dipole-dipole interactions \cite{Pich95} as Mn$^{2+}$ nominally has no orbital angular momentum.  The small anisotropy is consistent with {\MPS} having Heisenberg-like magnetism \cite{Wildes06}.  The anisotropy is very large for {\FPS}, which explains the Ising-like nature of its magnetism \cite{Joy92}.  The large anisotropy is likely due to a strong influence of Fe$^{2+}$ orbital angular momentum and has been previously discussed in the context of crystal field theory \cite{Joy92}.  

{\NPS} has a relatively small anisotropy, although the spin wave gap is relatively large due to the strength of its exchange parameters.  Orbital angular momentum will contribute to the anisotropy, although apparently not to the same degree as for {\FPS}.

As previously mentioned, the  anisotropy in {\NPS} has been the subject of debate.  The discrepancy may have less to do with the magnitude of $\Delta$ and more to do with the presence of the deep minimum in the spin wave dispersion at the C point in figure \ref{fig:Spagetti}.  This deep minimum suggests that the magnetic structure of {\NPS} is close to an instability.  {\NPS} has almost a hexagonal symmetry \cite{Ouvrard85}, and the $\left(0 1 0\right)$ and $\left(\frac{1}{2} \frac{1}{2} \frac{\overline{1}}{3} \right)$ reciprocal lattice points both have $Q \approx 0.6$ {\AA}$^{-1}$ and approximately map onto one another by rotating the reciprocal lattice by $60^{\circ}$.  Doing so would give a different magnetic structure with a propagation vector close to ${\bf{k}}_M = \left(\frac{1}{2} \frac{1}{2} \frac{1}{3}\right)$.  The possible instability may be coupled with the strong phonons found in the same energy range $E_{\Gamma}$ and $E_\text{C}$, as demonstrated in figure \ref{fig:Cuts} and in the appendix, leading to magnetostriction that distorts the magnetization if the crystal is physically constrained by, for example, glue \cite{Wildes15}.

With this in mind, some caution must be applied to the values determined from the experiments on powdered samples reported here.  The act of grinding the samples into powder may cause sufficient distortion to influence the magnetism.   Future efforts will focus on verifying the exchange parameters by measuring neutron scattering from single crystals.

In light of the apparent evolution shown in table \ref{tab:ExchangeComp}, it is interesting to determine the corresponding parameters for {\CPS} whose Co$^{2+}$ carry $S = 3/2$.  {\CPS} has an antiferromagnetic structure that is almost identical to {\NPS} and a N{\'e}el temperature similar to that of {\FPS} \cite{Wildes17}.  The paramagnetic susceptibility for {\CPS} is anisotropic in a manner similar to {\FPS} \cite{Joy92}, although the anisotropy is clearly different as the collinear axes for the ordered moments are almost orthogonal between the two compounds.  {\NPS}, however, has no apparent anisotropy in its paramagnetic susceptibility.  

The differences between the compounds are the result of removing electrons one at a time from a half-filled $d$ shell, introducing an orbital contribution to the magnetic moment in a relatively controlled manner.   This family of antiferromagnets will provide valuable information on spin-orbit interactions on a honeycomb lattice, particularly once the exchange parameters and anisotropy in {\CPS} are quantified, which is a topic of great current interest due to the search for Kitaev-Heisenberg systems \cite{Kitaev, Khaliullin}.

\section{Conclusions}
Neutron inelastic scattering has been used to determine the strengths of the magnetic exchange interactions and the anisotropy in {\NPS}.  The data were fitted using a Heisenberg Hamiltonian with a easy-axis single-ion anisotropy, and it was assumed that there was no magnetic exchange between the $ab$ planes.  The best results are shown in table \ref{tab:ExchangeComp}, showing that the first nearest-neighbour exchange is ferromagnetic, the second-nearest neighbour exchange is small,  and the third nearest-neighbour is very large and antiferromagnetic.  The measurements also establish the presence of a small anisotropy, giving rise to an energy gap of $\sim 7$ meV.  The analysis shows that a similar gap should be found in the Brillouin zone corner, suggesting that {\NPS} is close to a magnetic instability.

\section{Acknowledgements}

Experiments at the ISIS Neutron and Muon Source were supported by a beamtime allocation from the Science and Technology Facilities Council.  The authors wish to thank the {\ILL} for the use of their instruments.  ARW would like to thank Dr. B. F{\aa}k for stimulating discussions.  DL and ARW thank Prof. H. R{\o}nnow for financial support and for a critical reading of the manuscript.


\appendix
\section{\label{app:phonons} Phonon subtraction method}

The $Q-$dependence of the neutron scattering cross-section can be used to estimate the phonon contribution from the measured scattering.  The contribution can then be subtracted from the data and remaining signal can be considered to be purely magnetic.

The cross-section for phonons increases as  $\sim Q^2$ for small momentum transfers, eventually decreasing as $\sim\exp\left(-WQ^2\right)$ due to the Debye-Waller factor \cite{Squires},  The magnetic scattering, however, varies as the magnetic form factor squared which, for Ni$^{2+}$, decreases monotonically with increasing $Q$.  The scattering may thus be considered to be purely due to phonons at sufficiently large $Q$.

\begin{figure}
	\includegraphics[width=3.5in]{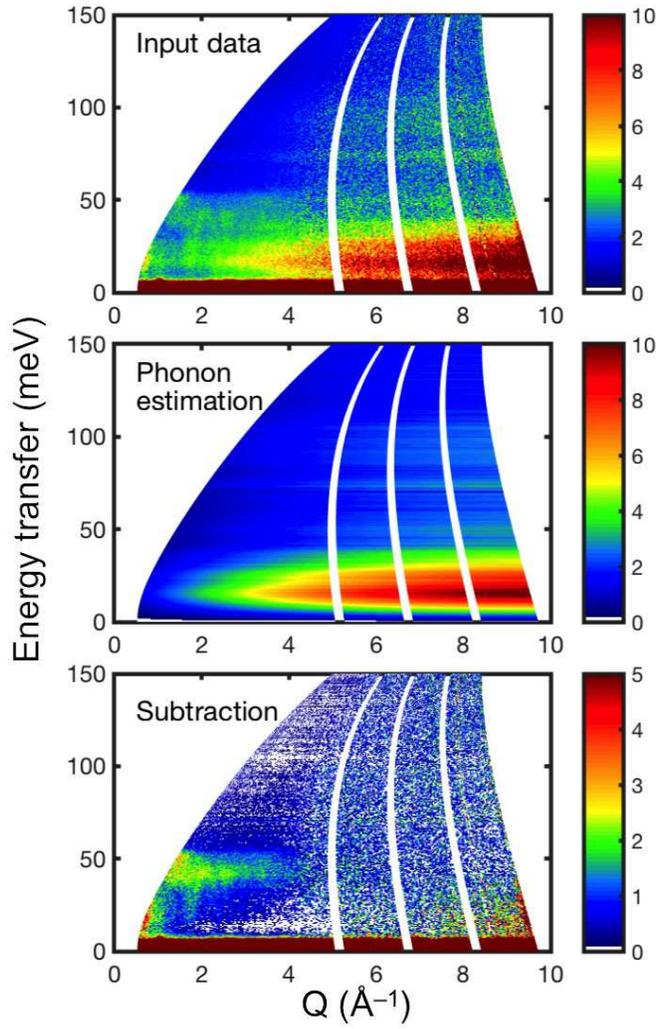}
	\caption{\label{fig:Phon} Neutron inelastic scattering from {\NPS} measured at 4 K on MAPS with $E_i = 200$ meV, the estimated phonon contribution to the data, and the MAPS data with the estimated contribution subtracted.}
\end{figure}

Figure \ref{fig:Phon} shows the neutron scattering data from MAPS over the full measured $Q-$range.  The magnetic inelastic scattering is visible at small $Q$, and it soon becomes swamped by the phonon contribution.    The phonon contribution was estimated in a two-step process.

The first step determined the $Q-$dependence of the phonon contribution.  Inspection of the data shows a reasonable density of phonon states from $70 \leq E \leq 120$ meV, which is greater than the maximum energy for the magnetic scattering.  These data were extracted and the intensities for each energy bin were fitted with the equation:
\begin{equation}
     I\left(Q,E\right) = Z_p\left(E\right)\left(p_1 + p_2Q^{p_4}\exp\left(-p_3Q^2\right)\right),
\label{eq:phonons}
\end{equation}
where $p_{1..3}$ were global fit parameters and $Z_p\left(E\right)$ is an amplitude for the phonons with energy $E$.   The exponent $p_4$ should nominally be equal to 2, however setting $p_4 = 1$ resulted in better fits to the data and this value was chosen for the subsequent data treatment.  The need to decrease $p_4$ may be understood as being the result of phonon multiple scattering in the sample.  The global parameters were found to be $p_1 = 0.146$, $p_2 = 0.597$ and $p_3 = 0.0047$.

The phonons over the entire energy range were assumed to have the same $Q-$dependence.  The second step was therefore to determine the values of $Z_p\left(E\right)$ for all $E$.  These were determined using the scattering at large $Q$.  The range $7 \leq Q \leq 8$ {\AA}$^{-1}$ was chosen for the MAPS data.  The detectors in this range were free from some spurious effects that were apparent at larger $Q$, and the magnetic form factor squared for Ni$^{2+}$ form factor for $Q > 7$ {\AA}$^{-1}$  is less than $0.002$ \cite{BrownFF}.  These data were extracted and fitted using equation \ref{eq:phonons} with $Z_p\left(E\right)$  being the only free parameter.

The method to estimate the phonon contribution becomes unreliable at low energies due to contamination from the elastic scattering.  Consequently, the method was only applied for energies above a minimum that was judged to be free from elastic contamination, which was chosen as 10 meV for the MAPS data.  The phonon contribution was assumed to vary linearly with $E$ below this energy, matching the gradient of the estimated phonon contribution for $10 \leq E \leq 13$ meV and becoming zero at the elastic line.

 The estimated phonon contribution for the MAPS data is shown in figure \ref{fig:Phon} along with the result of its subtraction from the experimental data.  There was some over-subtraction, particularly in the range of $E \sim 20$ meV which represented the peak in the phonon density of states.  The values of $Z_p\left(E\right)$ were therefore multiplied by 0.9 to reduce the over-subtraction.
 
 The data in the subtraction plot shown in figure \ref{fig:Phon} were used in the fitting.  A similar procedure was used for the MARI data.  The phonon-subtracted data from both instruments are shown in figure \ref{fig:Maps}.


\section*{References}

\begin{thebibliography}{99}
\bibitem{Brec} Brec R 1986 {\it Solid State Ionics} {\bf 22} 3
\bibitem{Grasso} Grasso V and Silipigni L 2002 {\it Riv. Nuov. Cim.} {\bf 25}  1
\bibitem{Ouvrard85} Ouvrard G, Brec R and Rouxel J 1985 {\it Mater. Res. Bull.} {\bf 20} 1181
\bibitem{Haines} Haines C R S, Coak M J, Lampronti G I, Liu C, Hamidov H, Wildes A R,  Daisenberger D, Nahai-Williamson P and S.S. Saxena S S  {\it{https://arxiv.org/abs/1801.10089}}
\bibitem{WangY} Wang Y, Ying J, Zhou Z, Sun J, Wen T, Zhou Y, Li N, Zhang Q, Han F, Xiao Y, Chow P,  Yang  W, Struzhkin V V, Zhao Y and H.-K. Mao 2018 {\it Nat. Comm.} {\bf 9} 1914
\bibitem{Park} Park J-G 2016 {\it J.Phys.: Condens. Matter} {\bf 28} 301001
\bibitem{Susner} Susner M A, Chyasnavichyus M, McGuire M A, Ganesh P and Maksymovych P 2017 {\it Adv. Mater.} {\bf 29} 1602852
\bibitem{WangF} Wang F, Shifa T A, Yu P, He P, Liu Y, Wang F, Wang  Z, Zhan X, Lou X, Xia F and He J 2018  {\it Adv. Func. Mater.} DOI:10.1002/adfm.201802151
\bibitem{Joy92} Joy P A and Vasudevan S 1992 {\it Phys. Rev. B} {\bf 46} 5425
\bibitem{Wildes98} Wildes A R, Roessli B, Lebech B and Godfrey K W 1998 {\it J. Phys.: Condens. Matter} {\bf 10} 6417
\bibitem{Wildes06} Wildes A R, R{\o}nnow H M, Roessli B, Harris M J and Godfrey K W 2006 {\it Phys. Rev. B} {\bf 74} 094422.
\bibitem{Wildes12} Wildes A R, Rule K C, Bewley R I, Enderle M and Hicks T J 2012 {\it J. Phys.: Condens. Matter} {\bf 24}  416004
\bibitem{Lancon} Lan{\c{c}}on D, Walker H C, Ressouche E, Ouladdiaf B, Rule K C, McIntyre G J, Hicks T J, R{\o}nnow H M and Wildes A R 2016 {\it Phys. Rev. B} {\bf 94} 214407
\bibitem{Wildes17} Wildes A R, Simonet V, Ressouche E, Ballou R and McIntyre G J 2017 {\it J. Phys.: Condens. Matter} {\bf 29} 455801
\bibitem{Wildes15} Wildes A R, Simonet V, Ressouche E, McIntyre G J, Avdeev M, Suard E, Kimber S A J, Lan\c{c}on D, Pepe G, Moubaraki B and Hicks T J 2015 {\it Phys. Rev. B} {\bf 92} 224408
\bibitem{Chandra} Chandrasekharan N and Vasudevan S 1994 {\it J. \ Phys.: \ Condens. \ Matter} {\bf 6} 4569
\bibitem{deJongh} Jongh L J and Miedema A R 2001 {\it Adv. Phys.} {\bf 50} 947
\bibitem{MARI} Taylor A D, Arai M, Bennington S M, Bowden Z A, Osborn R, Andersen K, Stirling W G, Nakane T, Yamada K and Welz D 1991 {\it KEK Report 90-25} {\bf 2} 705 (http://www.iaea.org/inis/collection/NCLCollectionStore/\_Public/22/090/22090885.pdf
\#page=75)
\bibitem{MAPS} Perring T G, Taylor A D, Osborn R, McK. Paul D, Boothroyd A T and G. Aeppli G 1994 {\it Proceedings of the 12th Meeting of the International Collaboration on Advanced Neutron Sources (ICANS XII), Cosener's House, Abingdon, Oxfordshire, UK, 24-28 May, 1993} I-60
\bibitem{BRISP} Formisano F, Francesco A D, Guarini E, Laloni A, Orecchini A, Petrillo C, Pilgrim W C, Russo D and Sacchetti F 2013 {\it J. Phys. Soc. Japan} {\bf SA028} (http://dx.doi.org/10.7566/JPSJS.82SA.SA028)
\bibitem{MANTID} Arnold O, Bilheux J C, Borreguero J M, Buts A, Campbell S I, Chapon L, Doucet M, Draper N, Ferraz Leal R, Gigg M A, Lynch V E, Markvardsen A, Mikkelson D J, Mikkelsone R L, Miller R, Palmen K, Parker P, Passos G, Perring T G, Peterson P F, Ren S, Reuter M A, Savici A T, Taylor J W, Taylor R J, Tolchenov R, Zhou W andJ. Zikovsky J 2014 {\it Nuc. Instrum. Meth. Phys. Res. Sect. A} {\bf 764} 156
\bibitem{LAMP} Richard D, Ferrand M and Kearley G J 1996 {\it J. Neutron Research} {\bf 4} 33
\bibitem{Wheeler} Wheeler E M, Coldea R, Wawrzy{\'n}ska E, S{\"o}rgel T, Jansen M, Koza M M, Taylor J, Adroguer P and Shannon N 2009 {\it Phys. Rev. B} {\bf 79} 104421
\bibitem{CHOP} {\it PyChop} http://docs.mantidproject.org/nightly/interfaces/PyChop.html (Accessed: 2018-08-1)
\bibitem{Rastelli} Rastelli E, Tassi A and Reatto L 1979 {\it Physica B} {\bf 97} 1
\bibitem{Fouet} Fouet J B, Sindzingre P and Lhuillier C 2001 {\it Eur. Phys. J. B} {\bf 20} 241
\bibitem{Kanamori} Kanamori J 1959 {\it J. Phys. Chem. Solids} {\bf 10} 87
\bibitem{Pich95} Pich C and F. Schwabl F 1995 {\it J. Magn. Magn. Mater.} {\bf 148} 30
\bibitem{Kitaev} Kitaev A 2006 {\it Ann. Phys.} {\bf 321} 2
\bibitem{Khaliullin} Khaliullin G 2005 {\it Prog. Theor. Phys. Suppl.} {\bf 160} 155
\bibitem{Squires} Squires G L 1996 {\it Introduction to the Theory of Thermal Neutron Scattering} Dover Publications, Mineola, New York
\bibitem{BrownFF} P. J. Brown P J 1995 {\it International Tables for Crystallography vol. C} ed. A. J. C. Wilson (Kluwer, Dordrecht) p. 391
\bibitem{VESTA} Momma K and Izumi F 2011 {\it J. Appl. Crystallogr.} {\bf 44} 1272









\end{thebibliography}

\end{document}